# Magnetic Torque in κ-(BETS)$_2$Mn[N(CN)$_2$]$_3$.

O.M.Vyaselev[1,*], S.V.Simonov[1], N.D.Kushch[2], W.Biberacher[3], and M.V.Kartsovnik[3]

[1]*Institute of Solid State Physics RAS, Chernogolovka, 142432, Russia*

[2]*Institute of Problems of Chemical Physics RAS, Chernogolovka, 142432, Russia*

[3]*Walther-Meissner-Institut, BAdW, Garching, B-85748, Germany*

**ABSTRACT**

Peculiarities observed in the field dependencies of the magnetic torque in κ-BETS-Mn measured at $T=1.5$K, $H=0-150$kOe, are interpreted from the viewpoint of two interacting spin subsystems, one associated with *d*-electron spins of $Mn^{2+}$ residing in the anion layer, and the other with itinerant π-electron spins that localize below the metal-insulator transition at $T_{MI} \approx 25$K, forming a long-range antiferromagnetic (AF) structure. The principal axes of the $Mn^{2+}$ spin subsystem have been defined. A model of AF π-spin arrangement is proposed that associates the observed 'kinks' in the field dependence of the torque with a spin-reorientation transition. One of the observed effects is ascribed to the π-*d* interaction between the two spin subsystems.

## I. Introduction

The radical cation salt κ-(BETS)$_2$Mn[N(CN)$_2$]$_3$, where BETS is $C_{10}S_4Se_4H_8$, bis-(ethylenedithio)tetraselenafulvalene), is a quasi-two-dimensional organic charge transfer complex consisting of sheets of organic donor BETS molecules responsible for the conductivity, sandwiched between insulating Mn[N(CN)$_2$]$_3^-$ anion layers [1]. This compound has been synthesized recently along with other akin materials like (BETS)$_2$FeX$_4$ (X=Cl, Br) [2,3], in a quest for hybrid multi-functional molecular materials combining conducting and magnetic properties in the same crystal lattice, potentially promising for microelectronics. κ-(BETS)$_2$Mn[N(CN)$_2$]$_3$ (hereafter referred to as κ-BETS-Mn) is a metal above $T_{MI} \approx 25$K and an insulator below this temperature [1] at ambient pressure. Under the applied pressure, $P$, the metal-insulator transition is removed and at $P > \approx 0.6$ kbar the system is a superconductor with a maximal $T_C \approx 5.8$K [4]. Measurements of magnetic properties of κ-BETS-Mn at ambient pressure [5, 6] have shown that the $Mn^{2+}$ spin subsystem, dominating the bulk magnetization of the sample, exhibits at $T > T_{MI}$ a Curie-Weiss-like behavior of a paramagnet with weak antiferromagnetic (AF) interactions. Below $T_{MI}$ the Curie-Weiss behavior is violated demonstrating a tendency to AF ordering. However, no long-range AF order of $Mn^{2+}$ spins has been detected below $T_{MI}$ [5,6], unlike (BETS)$_2$FeX$_4$ (X=Cl, Br) salts where the iron ions arrange antiferromagnetically in the insulating state [7-9]. Apparently, the key issue here is that in the latter case, $FeX_4^-$ anions are discrete units with a rectangular arrangement of magnetic $Fe^{3+}$ ions, while $Mn[N(CN)_2]_3^-$ anions of κ-BETS-Mn are of polymer type with a triangular lattice of $Mn^{2+}$ ions, which frustrates the AF ordering.

In contrast to electron spins of $Mn^{2+}$, those of π-electron (conducting) subsystem in κ-BETS-Mn do arrange antiferromagnetically in the insulating state, as has been unambiguously revealed in NMR experiments [10, 11]. A signature of AF order has been also detected in the magnetic torque measurements on κ-BETS-Mn [1,5]. In this paper we present a comprehensive set of magnetic torque data for κ-BETS-Mn and discuss the specific features caused by the presence of two interacting spin systems.

## II. Experimental.

The crystal structure of κ-BETS-Mn is monoclinic; the space group is $P2_1/c$ and the lattice constants at 200 K are: $a = 19.4723(9)$Å, $b = 8.4135(4)$Å, $c = 11.9608(7)$Å, $β = 92.315(4)°$, $V = 1957.94(17)$Å$^3$, and $ρ = 2.371$ g/cm$^3$, with two formula units per unit cell [4]. The conducting layers are formed by BETS dimers in the $(b,c)$ plane and sandwiched between the polymeric $Mn[N(CN)_2]_3^-$ anion layers in the $a$ direction. The crystal growth procedure and details of the structure have been reported elsewhere [1,4]. Results of the magnetization measurements have been reported previously [5].

The sample was a 40 μg thin-plate single crystal of ∼0.7×0.3×0.08 mm$^3$ size, with the largest dimensions along the conducting BETS layers [crystallographic $(b,c)$ plane]. Crystallographic directions of the crystal were X-ray defined. Magnetic torque was measured in fields up to 150 kOe with a homemade cantilever beam torquemeter described in Ref. 12. The cantilever was made of 50 μm thick as-rolled beryllium-copper foil. Technically the torque was determined from the capacitance change between the cantilever disc, to which the sample is attached, and the ground plate. The capacitance was measured using a tunable capacitive bridge. The maximum torque of the cantilever produced by the gravity force (in zero applied field) was 1.16 dyn·cm, the value used to convert the measured capacitance to the units of torque. The torquemeter was attached to a goniometer whose rotation axis was perpendicular to both the external magnetic field direction and the working plane of the cantilever. In this geometry, the component of the torque along the goniometer rotation axis is measured.

## III. Experimental Results.

Figure 1 exemplifies the field dependencies of the magnetic torque, τ, of the κ-BETS-Mn crystal measured at 1.5 K for the goniometer rotation axis aligned with directions $[0\bar{1}0]$ ($τ_b$,

Fig. 1a), [001] ($\tau_c$, Fig. 1b), $[0\bar{1}1]$ ($\tau_d$, Fig. 1c) and perpendicular to $[0\bar{1}1]$ ($\tau_{\perp d}$, Fig. 1d). Numbers to the right of the curves indicate the polar angle, θ, between the field direction and $a^*$, the direction perpendicular to the crystallographic (*bc*) plane.

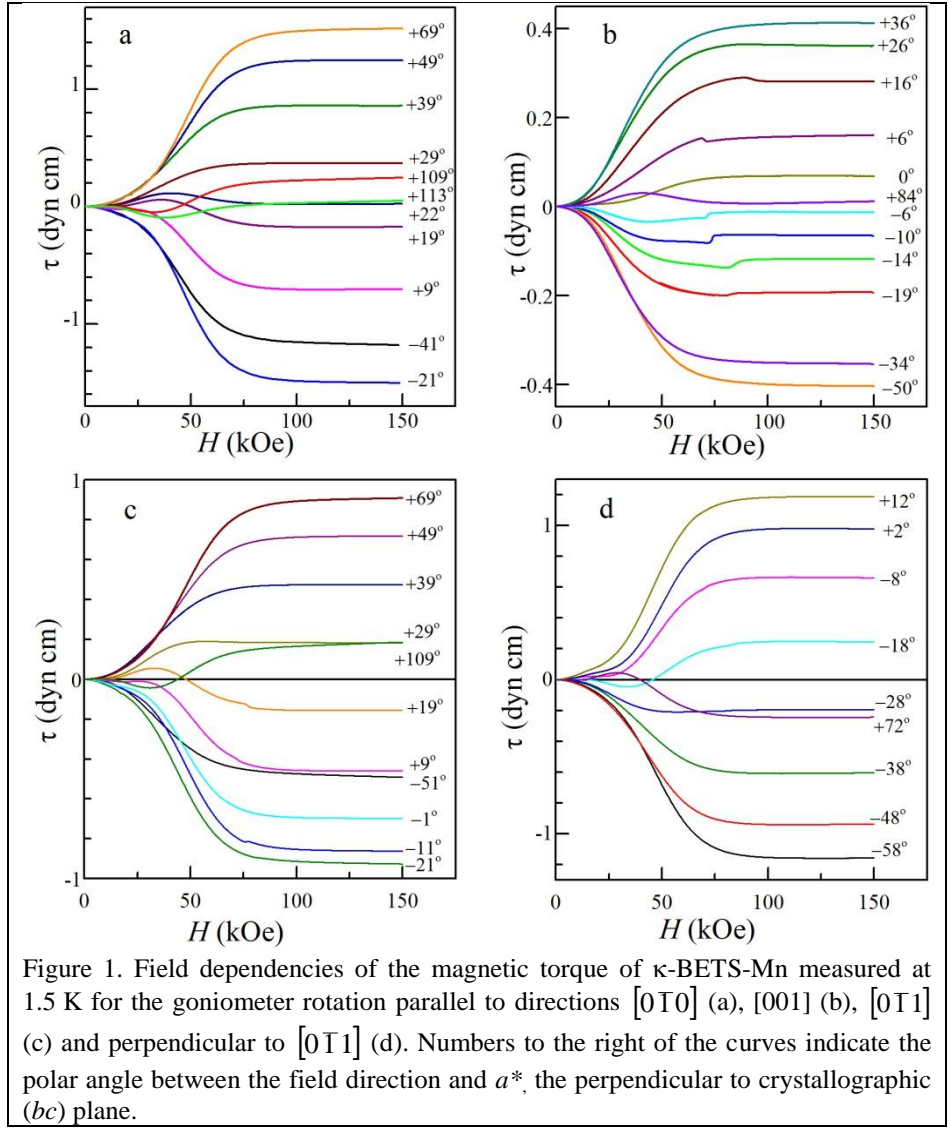

Figure 1. Field dependencies of the magnetic torque of κ-BETS-Mn measured at 1.5 K for the goniometer rotation parallel to directions $[0\bar{1}0]$ (a), [001] (b), $[0\bar{1}1]$ (c) and perpendicular to $[0\bar{1}1]$ (d). Numbers to the right of the curves indicate the polar angle between the field direction and $a^*$, the perpendicular to crystallographic (*bc*) plane.

There are several specific features visible in Fig. 1 that will be discussed below:

(i) At high field (*H* > 100 kOe) τ becomes constant;

(ii) For the angles where the torque is small at high field, τ is non-monotonic in the range 25–75 kOe;

(iii) At some angles $\tau_c$, $\tau_d$ and $\tau_{\perp d}$ (Figs. 1b, c, d, respectively) demonstrate a step-like feature ('kink') at fields 70–100 kOe. Fig. 2 represents the 'kinks' in more details. No such kinks have been detected for $\tau_b$ at any θ.

All the features listed above vanish as the temperature is raised above $T_{MI} \approx 25$ K. The observed non-monotonicity at intermediate fields, as well as the 'kinks' (Fig. 3) disappear, and the overall *H*-dependence of τ becomes a simple $H^2$, usual for an anisotropic paramagnet at $\mu_B H \ll k_B T$ (where $\mu_B$ is the Bohr magneton and $k_B$ is the Boltzmann constant). Important is that the directions of the field where the torque vanishes are exactly the same at high temperature as those at $T=1.5$K, $H=150$kOe. That means that the principal axes of the magnetization above and below $T_{MI}$ coincide. Therefore, the observed features should be associated with the low-temperature behavior of an anisotropic paramagnet and with the presence of the AF-ordered π-spin system.

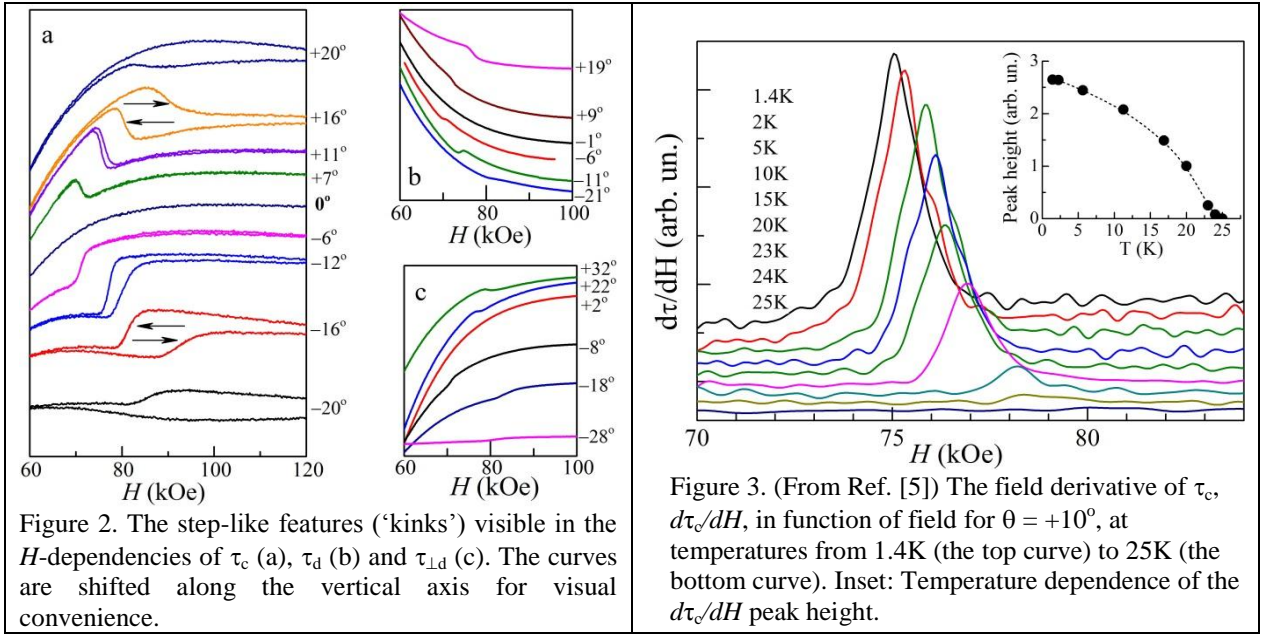

Figure 2. The step-like features ('kinks') visible in the $H$-dependencies of $\tau_c$ (a), $\tau_d$ (b) and $\tau_{\perp d}$ (c). The curves are shifted along the vertical axis for visual convenience.

Figure 3. (From Ref. [5]) The field derivative of $\tau_c$, $d\tau_c/dH$, in function of field for $\theta = +10°$, at temperatures from 1.4K (the top curve) to 25K (the bottom curve). Inset: Temperature dependence of the $d\tau_c/dH$ peak height.

## IV. Discussion

It was reported previously [5] that the torque in κ-BETS-Mn is dominated by Mn$^{2+}$ electron spins of the anion layer since it is several orders of magnitude bigger than in structurally similar magnetic ion-free organic salts. In turn, the observed 'kinks' have been related to the AF-ordered π-electron spin system and result from the spin-flop at corresponding fields. In what follows we characterize the phenomena dominated by each spin subsystem individually and the ramifications due to their interaction.

*General expressions for the magnetic torque.*

The magnetic torque is expressed as

$$\boldsymbol{\tau} = V\,\mathbf{M}\times\mathbf{B}, \qquad (1)$$

where $V$ is the volume of the sample, $\mathbf{M}$ is the sample magnetization and $\mathbf{B} = \mathbf{H}+4\pi\mathbf{M}$ is the magnetic field. Let us neglect for a while the ramifications due to the sample shape (that will be discussed below) and assume the sample is a sphere. In that case

$$\boldsymbol{\tau} = V\,\mathbf{M}\times(\mathbf{H}+4\pi\mathbf{M}) = V\,\mathbf{M}\times\mathbf{H}. \qquad (2)$$

Consider first the high-temperature, low-field limit, $\mu_B H \ll k_B T$. Assuming the field in the (*ab*) plane,

$$\mathbf{H} = H \begin{bmatrix} \cos\theta \\ \sin\theta \\ 0 \end{bmatrix} \qquad (3)$$

and the susceptibility tensor

$$\hat{\chi} = \begin{pmatrix} \chi_a & 0 & 0 \\ 0 & \chi_b & 0 \\ 0 & 0 & \chi_c \end{pmatrix}, \qquad (4)$$

one obtains the magnetization

$$\mathbf{M} = \hat{\chi} \cdot \mathbf{H} = H \begin{bmatrix} \chi_a \cos\theta \\ \chi_b \sin\theta \\ 0 \end{bmatrix}, \qquad (5)$$

and the torque $\boldsymbol{\tau} = V\,\mathbf{M}\times\mathbf{H} = [0, 0, \tau_c]$, where

$$\tau_c = \tfrac{1}{2}\, V H^2 (\chi_a - \chi_b)\sin 2\theta, \qquad (6)$$

which gives the observed quadratic in $H$ behavior of the torque at low fields/high temperatures.

In the high-field, low-temperature regime, the linearity of $M$ in $H$ of Eq. (5) is no more valid. At $\mu_B H \gg k_B T$ the magnetization of a paramagnet saturates to a constant value. In an anisotropic material the change in $H$ reduces to the change of the angle $\varphi$ between the magnetization vector and the field direction. Assuming that $a$ is the axis of easy magnetization ($\chi_a > \chi_b$) and for the field in the form (3), we write the magnetization vector in the form:

$$\mathbf{M} = M \begin{bmatrix} \cos(\theta-\varphi) \\ \sin(\theta-\varphi) \\ 0 \end{bmatrix}. \qquad (7)$$

The Zeeman energy of the system is

$$E_Z = -V\,\mathbf{M}\cdot\mathbf{H} = -V M H \cos\varphi, \qquad (8)$$

which minimizes at $\varphi = 0$ ($\mathbf{M}\|\mathbf{H}$). In the simplest uniaxial case, the magnetic anisotropy energy can be written in the form:

$$E_a = V K \sin^2(\theta-\varphi), \qquad (9)$$

where $K$ is the anisotropy constant. $E_a$ minimizes at $\varphi = \theta$, ($\mathbf{M}\,\|$ easy axis). The total energy, $E_Z + E_a$, minimizes when

$$d(E_Z + E_a)/d\varphi = V\,[MH\sin\varphi + K\sin 2(\theta-\varphi)] = 0. \qquad (10)$$

If the anisotropy term is small compared to the Zeeman term ($K \ll MH$), $\mathbf{M}$ tends to align with $\mathbf{H}$ and $\varphi \to 0$. Series-expanding Eq. (10) for $\varphi \to 0$, one obtains:

$$K\sin 2\theta + \varphi(HM + 2K\cos 2\theta) = 0, \qquad (11)$$

$$\varphi = -\frac{K\sin 2\theta}{HM + 2K\cos 2\theta}. \qquad (12)$$

As before, the torque $\boldsymbol{\tau} = V\mathbf{M}\times\mathbf{H} = [0, 0, \tau_c]$, where now

$$\tau_c = VMH\sin\varphi \approx VMH\varphi = -VK\sin 2\theta \frac{1}{1+\dfrac{2K\cos 2\theta}{HM}} \approx -VK\sin 2\theta. \qquad (13)$$

Expression (13) derived for the torque in the high-field limit explains the observed field-independent behavior of $\tau$ at high fields (Fig. 1).

*Locating the principal axes of the magnetization.*

Now we proceed to determining directions of the principal axes of the magnetization in κ-BETS-Mn. Figure 4 shows angular dependencies of the torque measured as $T=1.5$K, $H=150$kOe. The raw experimental data have been corrected for the torque resulted from the sample demagnetization, $\tau_{\text{dem}} = 2\pi VM^2\sin 2\theta$, as explained in the Appendix. For $M$ we use the maximum value 46.7 G of the saturated paramagnet with $L=0$, $S=5/2$, which seems to be a reasonable estimation according to the dc magnetometry data [5].

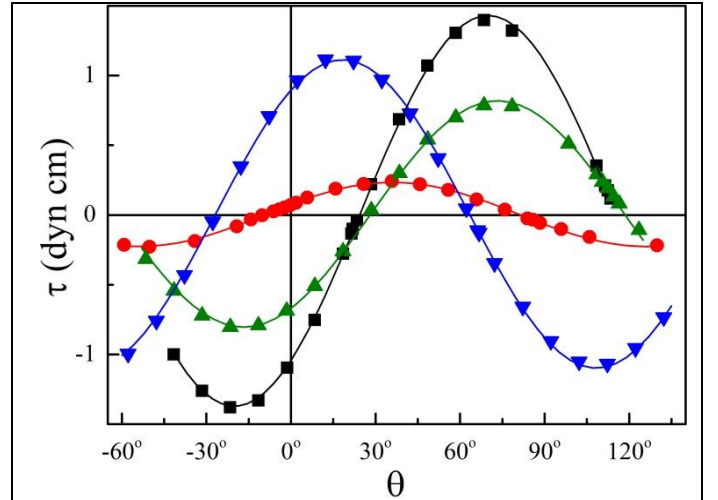

Figure 4. Angular dependencies of the torque in k-BETS-Mn at $T=1.5$K, $H=150$kOe for the field rotated around $[0\bar{1}0]$ (squares), $[001]$ (circles), $[0\bar{1}1]$ (up triangles) and perpendicular to $[0\bar{1}1]$ (down triangles). θ, the polar angle of the field direction is reckoned from $a^*$. Solid lines: fits to the data using Eq. (14).

All four curves nicely follow the $A\cdot\sin 2(\theta-\theta_0)$ dependence. However, for the practical purpose it is more convenient to present it in the form:

$$\tau = \alpha\cos 2\theta + \beta\sin 2\theta. \qquad (14)$$

The parameters α and β obtained from the fits to the data in Fig. 4 are listed in Table 1.

| Rotation axis | α | β | $\theta_0$ |
|---|---|---|---|
| $[0\bar{1}0]$ | −1.07 | 0.9 | 25° |
| $[001]$ | 0.071 | 0.218 | −9° |
| $[0\bar{1}1]$ | −0.678 | 0.447 | 28° |
| $\perp$ to $[0\bar{1}1]$ | 0.888 | 0.658 | −27° |

Table 1. Fit parameters to the torque data in Fig.4 according to Eq. (14).

In order to calculate the torque, we introduce the coordinate system $\{a,b,c\}$; $\mathbf{a} = [1,0,0]$ ($=\mathbf{a}^*$); $\mathbf{b} = [0,1,0]$; $\mathbf{c} = [0,0,1]$, where the $b$ and $c$ axes coincide with those of the crystal. The rotation axis vector is given by $\mathbf{R} = [0, -\cos\lambda, \sin\lambda]$, where λ is the angle between the rotation axis and the −**b** direction: for the rotation around $[0\bar{1}0]$ $\lambda=0$, around $[001]$ $\lambda=90°$, around $[0\bar{1}1]$ $\lambda=55°$ and around the perpendicular to $[0\bar{1}1]$ $\lambda=145°$.

At high field the linearity between **M** and **H** in the form of Eq. (5) is no more valid. In this case, in order to calculate the magnetization direction at a *given* value of $H$ we introduce the tensor

$$\hat{M} = \begin{pmatrix} d_{ab}+d_{ac}+d_{bc} & M_{ab} & M_{ac} \\ M_{ab} & -(d_{ab}-d_{ac})+d_{bc} & M_{ab} \\ M_{ac} & M_{bc} & d_{ab}-d_{ac}+d_{bc} \end{pmatrix}, \quad (15)$$

where $d_{ab} = (M_{aa}-M_{bb})/2$, $d_{ac}= (M_{aa}-M_{cc})/2$, $d_{bc}= (M_{bb}+M_{cc})/2$, so that $\mathbf{M} = \hat{M}\mathbf{h}$, where $\mathbf{h} = [\cos\theta, \sin\theta\sin\lambda, \sin\theta\cos\lambda]$ is the applied field unit vector. Then, using the expression for the torque given by Eq. (2), we obtain for the experimentally measured component of the torque on the rotation axis

$$\tau_r = H \{-\cos2\theta[M_{ac}\cos\lambda+M_{ab}\sin\lambda]+\sin2\theta[d_{ab}+d_{ac}-(d_{ab}-d_{ac})\cos2\lambda-M_{bc}\sin2\lambda)/2\}. \quad (16)$$

For the four rotation axes used in the experiment:

$$\tau_b\ (\lambda = 0) = H\{-M_{ac}\cos2\theta + d_{ac}\sin2\theta\}; \quad (17a)$$

$$\tau_c\ (\lambda = 90°) = H\{-M_{ab}\cos2\theta + d_{ab}\sin2\theta\}; \quad (17b)$$

$$\tau_d\ (\lambda=55°) = H\{-(0.819M_{ab}+0.574M_{ac})\cos2\theta+(0.67d_{ab}+0.33d_{ac}-0.47M_{bc})\sin2\theta\}; \quad (17c)$$

$$\tau_{\perp d}\ (\lambda=145°) = H\{-(-0.819M_{ac}+0.574M_{ab})\cos2\theta+(0.33d_{ab}+0.67d_{ac}+0.47M_{bc})\sin2\theta\}. \quad (17d)$$

In fact, a detailed inspection of the sample orientation for the rotation around *c*-axis revealed that the sample *c*-axis was slightly (by ~4°) misoriented from the direction of the rotation axis, and the correct value of $\lambda$ is 94°. In that case,

$$\tau_c\ (\lambda = 94°) = H\{-(M_{ab} - 0.07M_{ac})\cos2\theta + (d_{ab}+0.07M_{bc})\sin2\theta\} \quad (17e)$$

Equating the fit parameters α and β listed in Table 1 to the corresponding factors in Eqs. (17), one obtains the magnetization matrix:

$$\hat{M} = \frac{1}{H}\begin{pmatrix} 1.119+d_{bc} & 0.004 & 1.07 \\ 0.004 & 0.681+d_{bc} & -0.008 \\ 1.07 & -0.008 & -0.681+d_{bc} \end{pmatrix}, \quad (18)$$

Then the magnetization principal axes can be obtained as the eigenvectors of the tensor in Eq. (18): **x**=[0.907, 0.000, 0.422]; **y**=[0.002, 1, −0.004], **z**=[−0.422, 0.005, 0.907] for any arbitrary $d_{bc}$.

The (*xz*) plane of the magnetization eigenvectors practically (to within 0.3°) coincides with the (*ac*) plane of the crystal, while the **x** vector is directed at ≈25° from the *a\** direction in the (*ac*) plane. Very remarkably, this direction coincides with the orientation of the longitudinal axis of the BETS molecules in the crystal structure.

As it was mentioned above, at high temperature the directions of the field, where the torque zeroes, are exactly the same as those at $T=1.5$K, $H=150$kOe (Fig. 4). This means that the obtained orientations of the principal axes of the magnetization are inherent to the Mn$^{2+}$ spin system.

*Angular and Field dependence of the 'kinks'.*

The step-like features ('kinks') in $H$-dependencies of $\tau$ (Figs. 1 and 2) are observed at 70–100 kOe for the field rotated around [001] ($\tau_c$), [0$\bar{1}$1] ($\tau_d$) and a perpendicular to [0$\bar{1}$1] ($\tau_{\perp d}$), at angles $\sim 3^\circ < |\theta| < \sim 30^\circ$ from the $a^*$ direction. No such 'kinks' have been detected for $\tau_b$ (for the field rotated around [0$\bar{1}$0]) at any $\theta$. Figure 5 summarizes the angular behavior of the field where the 'kink' is observed, $H_{kink}$.

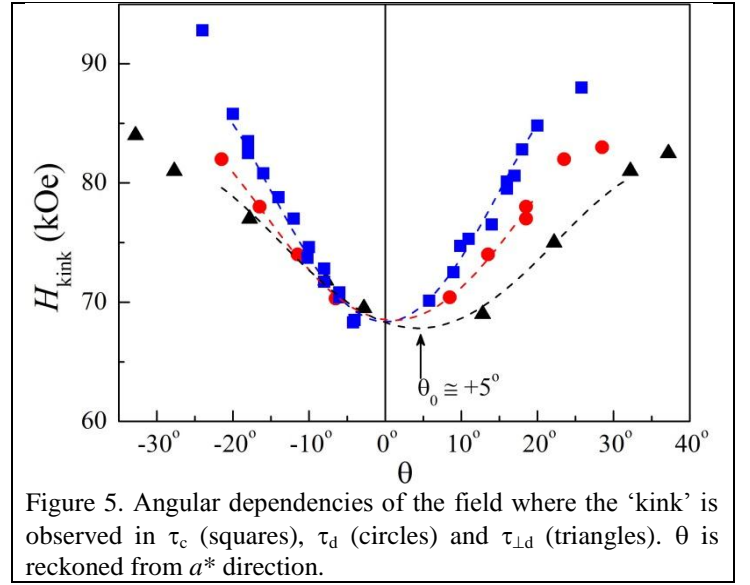

Figure 5. Angular dependencies of the field where the 'kink' is observed in $\tau_c$ (squares), $\tau_d$ (circles) and $\tau_{\perp d}$ (triangles). $\theta$ is reckoned from $a^*$ direction.

Therefore, the following conditions should be fulfilled in order to observe the 'kink':
- there must be a relatively large component of the field along $a^*$ and
- there must be a component of the field along [010] (the $b$-axis).

*Arrangement of the AF-ordered π-electron spins.*

A very detailed description of spin arrangement and field-induced spin reorientation (SR) in a Mott-insulating organic salt κ-(BEDT-TTF)$_2$Cu[N(CN)$_2$]Cl (BEDT-TTF = C$_{10}$S$_8$H$_8$, bis(ethylenedithio)-tetrathiafulvalene), which undergoes an AF transition below $T_N = 27$K and structurally is similar to k-BETS-Mn, has been given in Refs. [13,14]. The key concept is that in an AF system with a low symmetry of the crystal structure, the two magnetic sublattices $\mathbf{M}_1$ and $\mathbf{M}_2$ do not arrange strictly antiparallel along the easy axis but form a so-called canted ferromagnetic (CFM) order due to the Dzyaloshinskii-Moriya (DM) interaction [15, 16]. Following the notations of Ref. [14], we switch from the ($\mathbf{M}_1$, $\mathbf{M}_2$) to the ($\mathbf{M}_F$, $\mathbf{M}_S$) basis of ferromagnetic $\mathbf{M}_F = (\mathbf{M}_1 + \mathbf{M}_2)/2$ and staggered $\mathbf{M}_S = (\mathbf{M}_1 - \mathbf{M}_2)/2$ moments, as shown in Fig. 6.

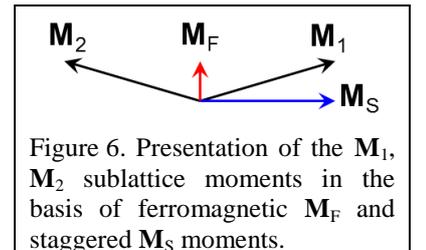

Figure 6. Presentation of the $\mathbf{M}_1$, $\mathbf{M}_2$ sublattice moments in the basis of ferromagnetic $\mathbf{M}_F$ and staggered $\mathbf{M}_S$ moments.

The free energy of the AF-ordered π-spin subsystem with the sublattice magnetizations outlined in Fig. 6, in the presence of the magnetic field is contributed from the Zeeman energy

$$E_z = -(\mathbf{M}_1 + \mathbf{M}_2) \cdot \mathbf{H} = -2\mathbf{M}_F \cdot \mathbf{H}, \tag{19}$$

the isotropic exchange energy

$$E_i = 2A(\mathbf{M}_1 \cdot \mathbf{M}_2) = 2A(M_F^2 - M_S^2), \tag{20}$$

the anisotropic exchange energy

$$E_a = 2K(\mathbf{M}_1 \cdot \mathbf{k})(\mathbf{M}_2 \cdot \mathbf{k}) = 2K\{(\mathbf{M}_F \cdot \mathbf{k})^2 - (\mathbf{M}_S \cdot \mathbf{k})^2\}, \tag{21}$$

and the DM term

$$E_{DM} = \mathbf{D} \times (\mathbf{M}_1 \times \mathbf{M}_2) = 2\mathbf{D} \times (\mathbf{M}_F \times \mathbf{M}_S), \tag{22}$$

where $A$ and $K$ are isotropic and anisotropic exchange constants, respectively, $\mathbf{k}$ the easy axis unit vector and $\mathbf{D}$ the DM vector. $E_z$ is minimized when $\mathbf{M}_F \parallel \mathbf{H}$, $E_i$ when $\mathbf{M}_1 = -\mathbf{M}_2$, or $|\mathbf{M}_F|=0$, $|\mathbf{M}_S|=M$ ($M$ is the magnitude of the electron spin moment of both sublattices): the spins minimize $E_i$ by aligning in an antiparallel orientation. $E_a$ is minimum when $\mathbf{M}_S \parallel \mathbf{k}$ because $E_a \ll E_i$, hence $\mathbf{M}_F \ll \mathbf{M}_S$ [13,17], and the effect of $E_{DM}$ is to arrange $\mathbf{M}_1$ and $\mathbf{M}_2$ perpendicular to each other, and both $\mathbf{M}_F$ and $\mathbf{M}_S$ perpendicular to $\mathbf{D}$. The ultimate spin orientation is determined by by a tradeoff between the four mechanisms.

The crystal structure of k-BETS-Mn assumes the mirror plane (*ac*). Symmetry considerations, thus, require $\mathbf{k}$ and $\mathbf{D}$ vectors in the (*ac*) plane and $\mathbf{M}_F$ along the *b* axis. Recent calculations [18] have shown that the preferable orientation of the $\mathbf{D}$ vector is the longitudinal axis of the BETS molecule, which is directed at ~21$^o$ from $a^*$ in the (*ac*) plane. The exact direction of $\mathbf{k}$ (AF easy axis) is currently unknown: there is the crystal anisotropy that specifies the $\mathbf{k}$ direction and the DM interaction which attempts to arrange $\mathbf{M}_S$ perpendicular to $\mathbf{D}$.

The zero-field arrangement of the AF sublattice moments is as follows: $\mathbf{M}_F$ is along the *b* axis and $\mathbf{M}_S$ in the (ac) plane at some angle with $\mathbf{D}$, as shown in the left panel in Fig.7. As the magnetic field is applied with a big enough component along $\mathbf{M}_S$ so that $E_z > E_a$, the orientation of $\mathbf{M}_F$ along the *b* axis becomes unfavorable and it switches to (or maybe close to) the direction of the external field. In turn, $\mathbf{M}_S$ switches to the direction perpendicular to both $\mathbf{M}_F$ and $\mathbf{D}$ (Fig.7, right panel). This is the so-called field-induced SR transition that causes the 'kinks' in the field dependence of the measured torque (Fig. 2). The angular dependencies of the transition field shown in Fig. 5 give a key to understand the orientation of $\mathbf{M}_S$ in the (*ac*) plane. One can notice that for the rotations around *c*-axis and [0$\bar{1}$1] direction, the 'kink' field is symmetric around θ=0, while for the rotation around the perpendicular to [0$\bar{1}$1] direction, which corresponds to rotation in the plane closest to (*ac*) plane, the hypothetic minimum in the angular dependence is shifted by some θ=5$^o$ from the $a^*$ direction. For this rotation plane the projection of the field vector with polar angle θ=5$^o$ to the (*ac*) plane makes an angle of ~–4° with $a^*$. Therefore, it is likely that $M_S$ at low fields is at some small angle from $a^*$ in the $(a, -c)$ quadrant.

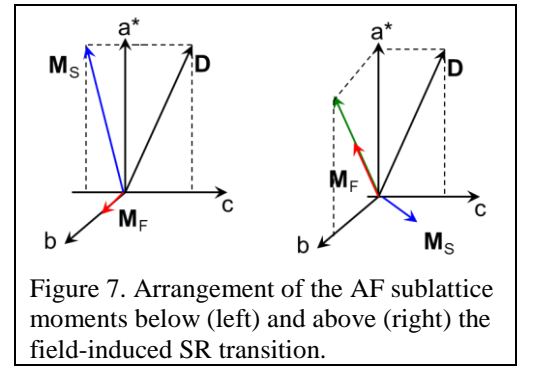

Figure 7. Arrangement of the AF sublattice moments below (left) and above (right) the field-induced SR transition.

The suggested model of the AF sublattice moments arrangement explains the existence of the 'kinks' in the field dependence of the measured torque, but does not explain why the 'kinks' are only observed when the external field has a non-zero $b$-axis component. One may doubt the existence of the SR transition when the field is along $a^*$, but recent NMR experiments have confirmed that it does exist [19]. The answer could be found in the crystal symmetry. Indeed, once $(ac)$ is the mirror plane, the zero-field orientation of $M_F$ along the $b$ and $-b$ directions are equally favorable, and there should be domains with either orientation of $M_F$. When the field directed along $a^*$ exceeds the critical value, the SR transition occurs but the sudden change in the torque caused by switching of $M_F$ from $b$ direction to the external field direction is compensated by the same process in the domains where $M_F$ is aligned with $-b$ at zero field. In contrast, a non-zero $b$-component of the applied field lifts this degeneracy, and the SR transition is revealed as a 'kink' in the torque $H$-dependence.

*Interaction between the two spin subsystems.*

One more peculiarity observed in the $H$-dependencies of the torque, is a non-monotonic behavior of $\tau$ in the field range 25–75 kOe for the directions of the field close to the magnetization principal axes, i.e. those where the high-field and high-temperature torque is zero. Apparently, this phenomenon is due to the $\pi$-$d$ interaction between the essentially paramagnetic $Mn^{2+}$ $d$-electron spin subsystem and the AF-ordered spin subsystem of $\pi$-electrons. An isolated $Mn^{2+}$ spin subsystem would produce a zero torque once the field is along any principal axis of the magnetization, since in that case the magnetization vector coincides with the field direction. However, at temperatures below $T_{MI} \approx 24K$ $\pi$-electron spins form a long-range AF order. As a consequence, the $d$-electron spin system of $Mn^{2+}$ acquires a certain magnetization due to the $\pi$-$d$ interaction even at $H=0$. The orientation of the zero-field magnetization of $Mn^{2+}$ depends on the details of the $\pi$-$d$ coupling, but it obviously does not coincide necessarily with the directions of the magnetization principal axes of the $Mn^{2+}$ subsystem. For a small field applied along a principal axis, the $Mn^{2+}$ magnetization does not coincide with the field direction resulting in a non-zero torque. However, as the field is increased above the value when the Zeeman energy is bigger than the $\pi$-$d$ coupling energy, the $Mn^{2+}$ subsystem magnetization returns to the direction of the field and the torque zeroes. One can see in Fig. 1 that this occurs at $H$~75kOe.

We need to note here that in the similar $\lambda$-(BETS)$_2$FeCl$_4$ compound the ground state of $Fe^{2+}$ spin subsystem has been reported a Néel-type antiferromagnet [7], as inferred from the dc magnetization data. However, later experiments have doubted this conclusion demonstrating the absence of AF correlations between $Fe^{2+}$ spins [20,21]. It is very likely, therefore, that the AF order of $Fe^{2+}$ spins observed in the magnetization measurements is entirely an effect of the π-d coupling that causes $Fe^{2+}$ moments to clone the AF arrangement of π-spins. In contrast, $Mn^{2+}$ spins do not show a prominent AF order [5]. As we mentioned above, the key issue here is that in the case of λ-$(BETS)_2FeCl_4$, $FeCl_4^{2-}$ anions are discrete units with a rectangular arrangement of magnetic $Fe^{2+}$ ions, while $Mn[N(CN)_2]_3^{2-}$ anions of κ-BETS-Mn are of polymer type with triangular lattice of $Mn^{2+}$ ions, which frustrates their AF-type arrangement. However, the π-d coupling is apparently the origin of the observed violation of the Curie-Weiss behavior of κ-BETS-Mn magnetization and the source for extra linewidth of $^1$H NMR below $T_{MI}$ [5,6].

## V.  Summary.

The anomalies observed in the field dependencies of the magnetic torque in κ-BETS-Mn measured at $T=1.5K$, $H=0-150kOe$, have been explained from the viewpoint of the two interacting spin subsystems, one associated with d-electron spins of $Mn^{2+}$ residing in the anion layer, and the other with π-electron spins that localize below the metal-insulator transition at $T_{MI}\approx25K$, forming a long-range AF structure. The principal axes of the $Mn^{2+}$ spin subsystem have been defined. A model of the AF π-spin arrangement has been proposed that associates the observed 'kinks' in the torque H-dependencies with a spin-reorientation transition. One of the observed effects is ascribed to the π-d interaction between the two spin subsystems.

## Appendix.

Consider an isotropic saturated ($\mu_B H \gg k_B T$) paramagnet in a shape of a thin oblate spheroid, placed in the external field $\mathbf{B}_{ex}$ directed at angle θ with the perpendicular to its surface:

$$\mathbf{B}_{ex} = B_{ex}[\cos\theta, \sin\theta, 0] \quad (A.1)$$

Once the material is assumed isotropic, the magnetization vector is parallel to $\mathbf{B}_{ex}$:

$$\mathbf{M} = M[\cos\theta, \sin\theta, 0] \quad (A.2)$$

The demagnetizing field is

$$\mathbf{B}_d = 4\pi\hat{n}\mathbf{M}, \quad (A.3)$$

where the demagnetizing factor

$$\hat{n} = \begin{pmatrix} n_\perp & 0 & 0 \\ 0 & n_\parallel & 0 \\ 0 & 0 & n_\parallel \end{pmatrix} \quad (A.4)$$

Then one immediately obtains $\boldsymbol{\tau} = V\mathbf{M}\times\mathbf{B} = V\mathbf{M}\times(\mathbf{B}_{ex}-\mathbf{B}_d) = [0, 0, \tau_c]$ where

$$\tau_c = 2\pi(n_\perp - n_\parallel)VM^2\sin2\theta. \quad (A.5)$$

In the case of a very thin sample $n_\perp = 1$, $n_\parallel = 0$ and $\tau_c = 2\pi VM^2\sin2\theta$.

**Acknowledgements.**

The work was supported by the Russian Foundation for Basic Research (Project No. 13-02-00350) and Deutsche Forschungsgemeinschaft (DFG Project KA 1652/4-1).